\documentclass[fleqn,usenatbib]{mnras}

\usepackage{newtxtext,newtxmath}

\usepackage[T1]{fontenc}
\usepackage{ae,aecompl}

\usepackage{graphicx}    
\usepackage{amsmath}    
\usepackage{amssymb}    
\usepackage{tabularx}   

\newcommand{\Mzeta}{$\rm{ \zeta\ }$}
\newcommand{\MTvir}{$\rm{ T_{vir}\ }$}
\newcommand{\MLX}{$\rm{ L_X/SFR\ }$}
\newcommand{\MEo}{$\rm{ E_0\ }$}
\newcommand{\code}[1]{\textsc{\small #1}}
\defcitealias{Greig2018}{GM18}
\newcommand{\quotes}[1]{``#1''}

\title[Learning EoR]{Deep learning from 21-cm tomography of the Cosmic Dawn and Reionization}

\author[N. J. F. Gillet et al.]{
Nicolas Gillet$^{1}$,\thanks{E-mail: nicolas.gillet@sns.it}
Andrei Mesinger$^{1}$,
Bradley Greig$^{2,3}$,
Adrian Liu$^{\dagger,4,5}$,
Graziano Ucci$^{1}$
\\
$^{1}$Scuola Normale Superiore, Piazza dei Cavalieri 7, I-56126 Pisa, Italy\\
$^{2}$ARC Centre of Excellence for All-Sky Astrophysics in 3 Dimensions (ASTRO 3D), University of Melbourne, VIC 3010, Australia\\
$^{3}$School of Physics, University of Melbourne, Parkville, VIC 3010, Australia\\
$^{4}$Department of Astronomy and Radio Astronomy Laboratory, University of California Berkeley, Berkeley, CA 94720, USA\\
$^{5}$Department of Physics and McGill Space Institute, McGill University, Montreal QC H3A 2T8, Canada\\
$\dagger$ Hubble Fellow
}

\date{Accepted XXX. Received YYY; in original form ZZZ}

\pubyear{2018}

\begin{document}
\label{firstpage}
\pagerange{\pageref{firstpage}--\pageref{lastpage}}
\maketitle

\begin{abstract}
The 21-cm power spectrum (PS) has been shown to be a powerful discriminant of reionization and cosmic dawn astrophysical parameters. However, the 21-cm tomographic signal is highly non-Gaussian.  Therefore there is additional information which is wasted if only the PS is used for parameter recovery. Here we showcase astrophysical parameter recovery directly from 21-cm images, using deep learning with convolutional neural networks (CNN).  Using a database of 2D images taken from 10,000 21-cm light-cones (each generated from different cosmological initial conditions), we show that a CNN is able to recover parameters describing the first galaxies: (i) \MTvir, their minimum host halo virial temperatures (or masses) capable of hosting efficient star formation; (ii) \Mzeta, their typical ionizing efficiencies; (iii) \MLX, their typical soft-band X-ray luminosity to star formation rate; and (iv) \MEo, the minimum X-ray energy capable of escaping the galaxy into the IGM.  For most of their allowed ranges, log \MTvir and log \MLX are recovered with $< 1 \%$ uncertainty, while \Mzeta and \MEo are recovered with $\sim 10\%$ uncertainty.  Our results are roughly comparable to the accuracy obtained from Monte Carlo Markov Chain sampling of the PS with \code{21CMMC} for the two mock observations analyzed previously, although we caution that we do not yet include noise and foreground contaminants in this proof-of-concept study.  
\end{abstract}

\begin{keywords}
cosmology: theory -- dark ages, reionization, first stars -- diffuse radiation -- early Universe -- galaxies: high-redshift -- intergalactic medium
\end{keywords}


\section{Introduction}
\label{sec:Introduction}
The cosmic dawn (CD) of the first galaxies and subsequent reionization of the Universe remain among the most compelling yet elusive cosmological epochs.  Little is currently known beyond approximately when the bulk of reionization occurred.  The properties of the unseen first galaxies and intergalactic medium (IGM) structures thought to govern this cosmic milestone, remain unknown.

Fortunately, the field is set to undergo a Big Data revolution, driven by interferometric observations of the cosmic 21-cm signal. Corresponding to the spin-flip transition of neutral hydrogen, the 21-cm line is sensitive to the thermal and ionization state of the IGM, making it an ideal probe of CD and the epoch of reionization (EoR).  Current interferometers, such as the Low Frequency Array (LOFAR; \citealt{Haarlem2013, Yatawatta2013}), the Murchison Wide Field Array (MWA; \citealt{Tingay2013}), and the Precision Array for Probing the Epoch of Reionization (PAPER; \citealt{Parsons2010}), are hoping for a statistical detection of the EoR. The upcoming Hydrogen Epoch of Reionization Array (HERA; \citealt{Deboer2017}) will go beyond that, capturing the fluctuations of the signal over a large range of scales and redshifts, allowing us to tightly constrain galaxy properties (e.g. \citealt{Greig2018}).  Eventually, the Square Kilometer Array (SKA; \citealt{Mellema, Koopmans2015}) will allow us to do high signal-to-noise (S/N) imaging of the EoR and CD, providing a 3D map of the first billion years of our Universe.

The timing and patterns of the signal encode the star formation histories, as well as the UV and X-ray properties of the first galaxies. The challenge is in interpreting the signal, in order to learn these properties.  Early work showed general qualitative trends.  For example, if the EoR is driven by rare, bright galaxies, the resulting 21-cm power would be larger than if it were driven by abundant, faint galaxies (e.g. \citealt{Furlanetto2004, McQuinn2007,Iliev2012}).  Abundant absorbers in the IGM (so-called Lyman limit systems; LLSs) would suppress the large-scale power (e.g. \citealt{McQuinn2007,Sobacchi2015}).  Hard X-ray sources would heat the IGM more uniformly, compared to soft X-ray sources, thus decreasing the available contrast in 21-cm images of the CD (e.g. \citealt{Pacucci2014,Fialkov2017}). 

These trends can now be quantified in detail,  given the advent of efficient Monte Carlo Markov Chain (MCMC) samplers of 21-cm simulations, such as \code{21CMMC}\footnote{ https://github.com/BradGreig/21CMMC} (\citealt{Greig2015, Greig2017, Greig2018} hereafter referred to as \citetalias{Greig2018}).  For a given parametrization of astrophysics, \code{21CMMC} computes the parameter constraints available from upcoming 21-cm observations. However, a choice must be made which summary statistic is used in computing the likelihood (i.e. to quantify the similarity between a prediction based on a particular parameter set and the observation).  A simple and popular choice of likelihood statistic is the 21-cm power spectrum (PS).  Indeed the PS was shown to be a powerful discriminant of reionization and cosmic dawn astrophysics (e.g. \citetalias{Greig2018}).

However, the 21-cm signal is highly non-Gaussian, as various radiation fields, driven by biased sources, induce complicated correlations in the ionization and thermal state of the gas (e.g. \citealt{Barkana2008, Bharadwaj2005,Zahn2007,Shimabukuro2017,Majumdar2018,Watkinson2018}).  Therefore there is much additional information which is wasted if only the PS is used for parameter recovery (see for example Fig. 1 in \citealt{Mellema2014}). Motivated by this, {\it here we showcase astrophysical parameter recovery directly from 21-cm images, using machine learning (ML)}.

Machine learning is powerful because it allows a model to adapt and learn complex relationships in data, without requiring the user to {\it a priori} specify functional forms.  ML is becoming popular in various fields of astronomy (e.g \citealt{Kamdar2016, Kamdar2016a, Ucci2017, Ucci2018,  Parks2017, Schaefer2017, Rodriguez2018, Gupta2018}).
Recently, ML was also applied to the 21-cm signal, by creating an emulator to replace more expensive simulation codes \citep{Kern2017, Schmit2018} or performing astrophysical parameter recovery with the PS statistic \citep{Shimabukuro2017}. 

In this study, we will use a Convolutional Neural Network (CNN) which is a ML technique designed to work on images. CNNs are widely used today, most famously for facial recognition and image classification.  Applying a CNN directly on a 21-cm image allows the network to adaptively chose summary statistics when performing parameter inference, rather than {\it a priori} specifying the summary statistic (such as a PS).  Thus, the CNN can implicitly take advantage of non-Gaussian information in the images.

This work is organized as follows. In \S \ref{sec:21-cm images} we present our cosmological 21-cm simulations and the resulting database of images. Then in \S \ref{sec:Convolutional Neural Network} we describe what is a CNN. Finally, in \S \ref{sec:Results: Parameter inference with a CNN} we quantify the performance of the CNN in astrophysical parameter inference from 21-cm images. Unless stated otherwise, we quote all quantities in co-moving units and adopt the cosmological parameters: ($\Omega_\Lambda$, $\Omega_{\rm M}$, $\Omega_b$, $n$, $\sigma_8$, $H_0$) = (0.69, 0.31, 0.048, 0.97, 0.81, 68 km s$^{-1}$ Mpc$^{-1}$), consistent with recent results from the Planck mission \citep{PlanckCollaboration2015}.

\section{21-cm images}
\label{sec:21-cm images}

We start by briefly describing the simulations used to create the 21-cm images, before proceeding to discuss the CNN techniques.  For more details about the simulation set-up, we refer the interested readers to \citetalias{Greig2018} and references therein.

\subsection{Database of simulated light-cones}
\label{sec:Database of simulated light-cones}

We simulate our 21-cm light-cones (LC) using the public code, \code{21cmFAST}\footnote{https://github.com/andreimesinger/21cmFAST} \citep{Mesinger2007,Mesinger2011}. In order to do a direct comparison to the parameter recovery using the PS with \code{21CMMC}, we use the same version of the code and free parameters from \citetalias{Greig2018}. Specifically, we vary four astrophysical parameters found to have the strongest impact on the 21-cm signal (and thus having the tightest parameter constraints):
    \begin{itemize}
    
    \item \Mzeta: the UV ionizing efficiency of galaxies.  This efficiency can be expressed as: 
    \begin{equation}
    \zeta = 30 \left( \frac{f_{esc}}{0.1} \right) \left( \frac{f_{*}}{0.05} \right) \left( \frac{N_{\gamma/b}}{4000} \right) \left( \frac{1.5}{1+n_{rec}} \right)
    \end{equation}
where, $\rm{f_{esc}}$ is the fraction of ionizing photons escaping into
the IGM, $\rm{f_{\ast}}$ is the fraction of galactic gas in stars, $\rm{N_{\gamma/b}}$ is the number of ionizing photons produced per baryon in stars and $\rm{n_{rec}}$ is the typical number of times a hydrogen atom recombines.  \Mzeta primarily controls the timing of the epoch of reionization (EoR). As in \citetalias{Greig2018} we consider the range $\rm{\zeta \in [10-250] }$.
    
    \item \MLX: the soft X-ray emissivity (below 2 keV)\footnote{Harder photons have mean free paths longer than the Hubble length at the redshifts of interest, and thus do not contribute to heating the IGM (e.g. \citealt{McQuinn2012,Das2017}).} per unit of star formation rate escaping the galaxy.  X-rays are responsible for heating the neutral IGM during the CD, during the so-called Epoch of Heating (EoH).  \MLX primarily impacts the timing of the EoH, with lower values delaying heating and thus increasing the strength of the absorption signal. As in \citetalias{Greig2018} we consider the range $\rm{ log_{10}(L_{X<2keV}/SFR) \in [38-42] ~ erg ~ s^{-1} ~ M_{\odot}^{-1} ~ yr }$.
  
      \item \MTvir: the minimum virial temperature of halos capable of hosting star forming galaxies. Star formation is suppressed for halos below this threshold, due to feedback and/or inefficient gas cooling. Since star formation governs all epochs of the 21-cm signal, \MTvir affects the timing of the cosmic milestones, e.g. EoR, EoH, and Wouthuysen-Field (WF) coupling \citep{Wouthuysen1952, Field1958}, thus affecting the entire signal.  Moreover, \MTvir determines the bias of the typical galaxy population and the resulting radiation fields. As in \citetalias{Greig2018} we consider the  range $ \log_{10} \rm{ (T_{vir} / 1 K ) \in [4-6] } $.

    \item \MEo: the X-ray energy threshold for self-absorption by the galaxy.  X-ray photons below this energy are absorbed by the interstellar medium (ISM) of the host galaxies. \MEo determines the hardness of the X-ray SED escaping the first galaxies.  Since the absorption cross-section is a strong function of energy, \MEo governs how homogeneous and efficient is the X-ray heating.  We consider the range, $\rm{ E_{0} \in [0.1-1.5]\ keV }$, equivalent to an average \ion{H}{I} column density of $\rm{ log_{10}(N_{\ion{H}{I}}) \in [19.3 - 23.0]\ cm^{-2} }$.
\end{itemize}

Two additional parameters were studied in \citetalias{Greig2018}: $\rm{R_{MFP}}$, the maximum ionizing photon horizon within ionized regions, and $\rm{\alpha_{X}}$, the X-ray spectral energy index. Since the authors found that these parameters have a comparably small impact on the 21-cm signal (c.f. Fig. 1 of \citealt{Greig2017}), in this work we fix them to $\rm{ R_{MFP}=15\ Mpc }$ and $\rm{\alpha_{X}=1}$.
 
We generate a database of 10,000 21-cm light-cones by randomly sampling the 4 astrophysical parameters, uniformly over the ranges quoted above.  We stress that {\it each light-cone is generated from an independent realization of the initial Gaussian random field}.  This is very important for machine learning, as otherwise, the network can adapt to features at a specific position in the image, resulting in spuriously good results.

The transverse faces of the resulting light-cones\footnote{To be precise, they are actually light-{\it cuboids}, though we stick with {\it cone} as per convention.} are 300$\times$300 Mpc, while they extend from $\rm{z=6}$--30 in the parallel direction (line of sight direction).  The resolution of the light-cones is 1.5 Mpc, corresponding to the dimensions of 200$\times$200$\times$2200.

\subsection{Reducing the light-cones to 2D images}
\label{sec:Reducing the light-cones to 2D images}

\begin{figure*}
\includegraphics[width=\textwidth]{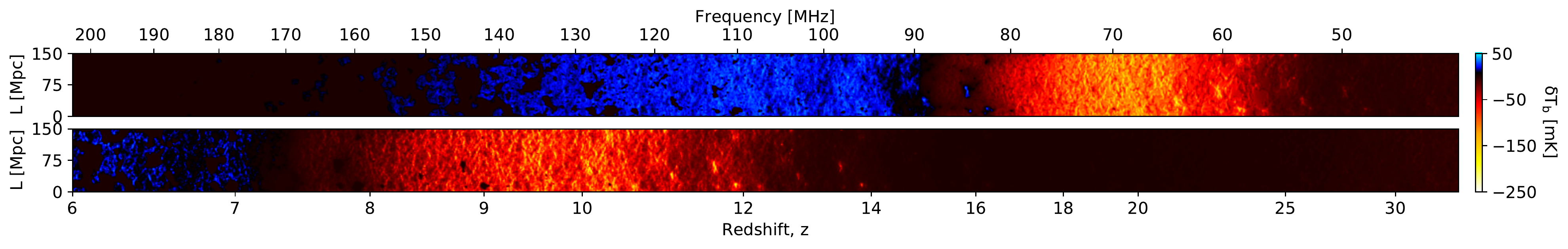}
\caption{ Example of the reduced, 2D 21-cm light-cone images used in the learning phase of the CNN.  The top/bottom panel corresponds to the FAINT GALAXIES / BRIGHT GALAXIES mock observation in \citetalias{Greig2018} (see their Figure 1). }
\label{fig:LC_faint_bright}
\end{figure*}

Convolutional Neural Networks require many iterations and tests in order to select the best architecture and tune the various free hyper-parameters (see the next section for details).  Performing such optimization with large data sets can be computationally expensive.  For reference, our fiducial CNN described in the next section would take of order a month to train on the full database of 3D light-cones, using 30 CPUs.  Even perfunctory optimization of the CNN architecture would require tens of these training runs, which is not currently computationally practical.

\begin{figure}
\includegraphics[width=\columnwidth]{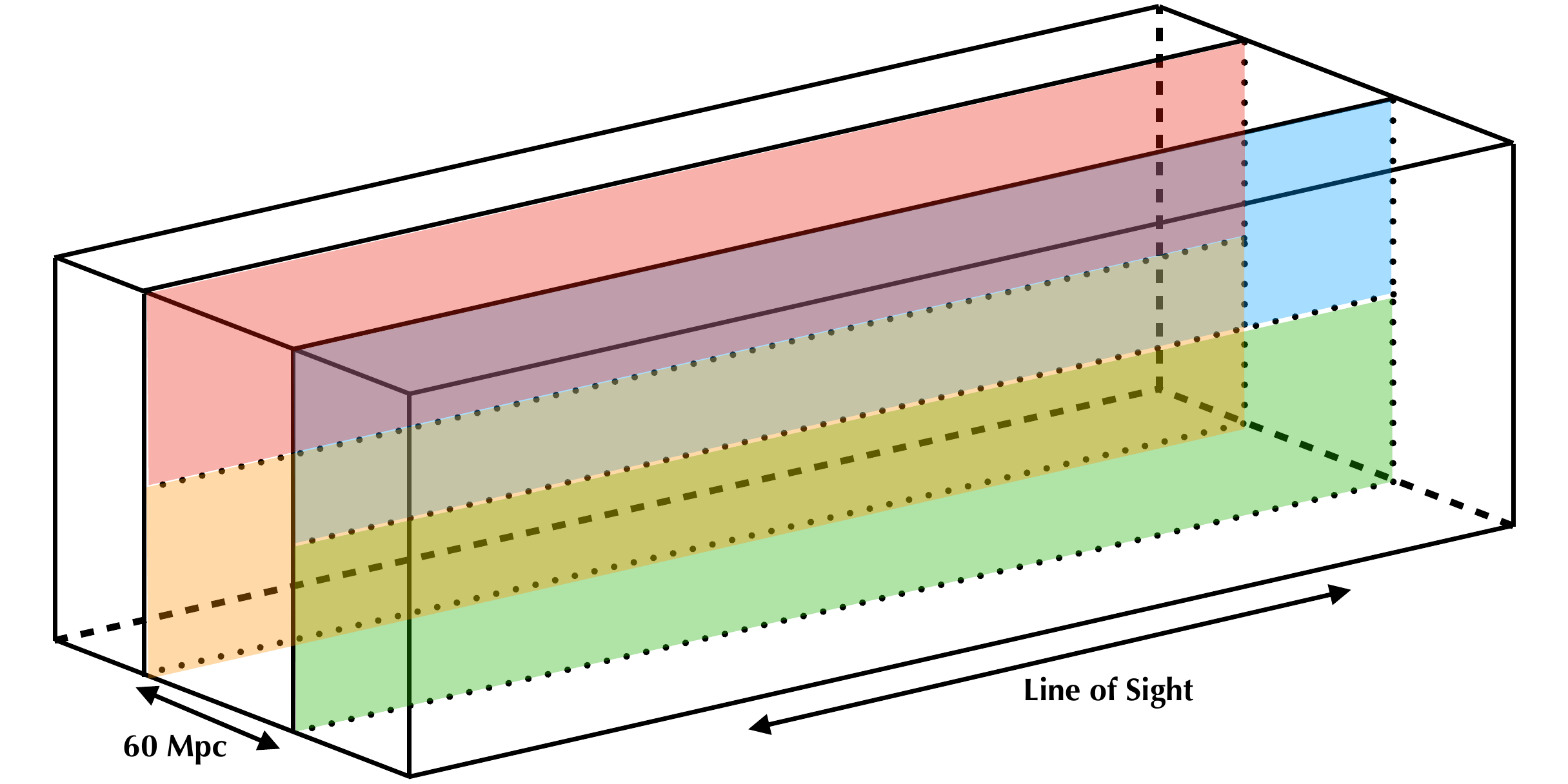}
\caption{ Scheme illustrating the slicing of light-cones.  Two slices along the line-of-sight are represented, separated by 60 Mpc.  Each slice is again divided in two along the line of sight, leading to four "light-cone images" on the scheme: blue, green, red and orange.  In practice five slice are taken separated by 60 Mpc, producing 10 images per light-cone. }
\label{fig:scheme_slicing}
\end{figure}

For this reason, we instead train our CNN on 2D slices through the light-cones (see Fig. \ref{fig:LC_faint_bright}).  The slicing is illustrated by the scheme in Fig. \ref{fig:scheme_slicing}.  Specifically, we take 5 slices along the line-of-sight (LOS) axis for each light cone, separated by 60 Mpc, thus preserving the redshift evolution.  Moreover, we split those slices in two, so that the training set images each correspond to half of the transverse volume (150 Mpc), which should preserve most of the large-scale structures (e.g. \citealt{Iliev2014}). The resulting 2D slices are 100$\times$2200 pixels. The resulting 10 slices per LC represent a 40-fold compression from the full 3D light cuboid of 200$\times$200$\times$2200. Our fiducial CNN takes under 5 days to train on the resulting set of 2D images.

Training the CNN on this reduced data is a fairly conservative choice: keeping more data could result in better parameter recovery.  However, for this proof-of-concept, which requires many iterations to tune hyper-parameters, we opt for computational efficiency.  In the future, we plan to extend the CNN to train on the full 3D light-cones, taking advantage of the speed-ups available with GPU-optimized ML toolkits.

\section{Convolutional Neural Network}
\label{sec:Convolutional Neural Network}

Here we go into detail about what is a CNN, and how we use it to perform parameter inference.  We start with some general discussion on ML, before focusing on our specific application of a CNN on 21-cm images.

\subsection{ Introduction to Machine Learning and Neural Networks}
\label{sec:Introduction to Machine Learning and Neural Networks}

ML allows the modeling of physical processes without having to specify functional forms.  It is composed of free parameters that will be adapted based on some data i.e. the relation between inputs and outputs will be learned without {\it a priori} parametrization. An example could be predicting the star-formation rates of galaxies (output), based on their DM halo properties (input).  This learning is an iterative process, where the model adjusts itself depending on the difference between its prediction and the \quotes{truth}. The learning phase requires a training set for which the \quotes{true} answer is known (e.g. the outputs of hydrodynamic galaxy formation simulations for the example above).

A Neural Network (NN) is a type of ML. The base block of a NN is a neuron.  A neuron takes an input array {\it x} and perform a linear combination with an array of trainable weights {\it w} plus a bias factor {\it b} to compute the net-input {\it s}.  In other words, a given neuron $j$ produces the following output $s^j$ from an input vector $x_i$:

\begin{equation}
s^j = \sum_{i=input} w^j_i x_i + b^j.
\end{equation}

Neurons are organized in layers.  All neurons in a given layer see the same input vector  $x$, but each neuron has its own independent weights vector $w^j$.  The resulting $s^j$ vector outputted by a given layer is then used to make the input for the successive layer (which could be comprised by a different number of neurons).  Several layers of neurons then collectively form the NN.

The linear combinations of each layer can be used to reproduce arbitrary, non-linear functions, using the so-called activation function $\phi(s)$.  It is applied to the output of each neuron, in order to quantify if it is significant or not. In essence, the activation function serves as an \quotes{on/off} switch,  allowing the NN to ignore the outputs of irrelevant neurons. Typical activation functions are sigmoid, hyperbolic-tangent or Rectified Linear Units (ReLU).  The choice of the activation function is a free hyper-parameter of the network.  Here we choose the ReLU activation function \citep{Nair2010}: $\rm{ \phi(s)=s\ if\ s>0,\ otherwise\ 0}$.  The neurons of the final layer which produces the desired output do not have activation functions.

NNs are very efficient when applied to relatively small input arrays.  However, since the number of weights which need to be learned depends on the size of the input, learning can become very slow for large or high-dimensional data.  More specific types of NNs have been developed to account for this.

\begin{figure}
\includegraphics[width=\columnwidth]{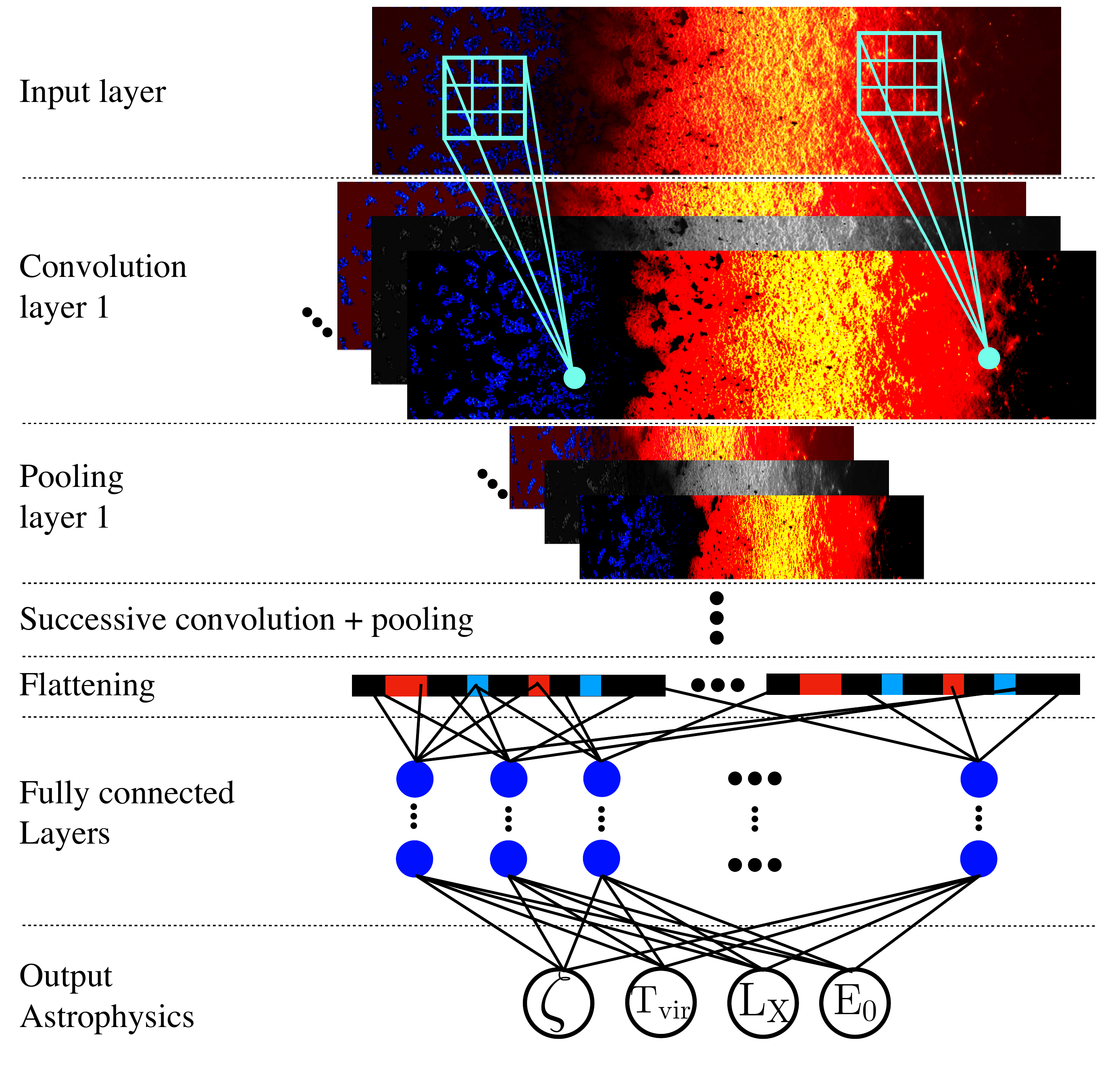}
\caption{ A schematic of a CNN.  The first part of a CNN takes an image and performs a series of convolutions with square filters (here shown with a 3x3 matrix), with each convolution having its own adjustable weights.  The results from these filters are fed to neurons (one per pixel; here represented by blue circles) which have their own activation function.  This data is further reduced by pooling (down-sampling) and then again performing convolution and pooling using as input the output of the preceding layer.  The second part consists of a classical NN operating on a flattened (1D) image resulting from the convolution and pooling layers. At the flattening stage, the colors illustrate the propagated information from the EoH (red) and EoR (blue). The outputs of the network are astrophysical parameters, shown at the bottom. Note that the colors and values used here are purely illustrative; examples of actual inputs/outputs of our convolution layers can be found in the appendix.  For clarity, only one convolution and pooling layer are shown, and the filters of the convolution layer are illustrated by a 3x3 matrix, instead of the 10$\times$10 matrix used in our CNN.
}
\label{fig:CNN_scheme}
\end{figure}

\subsection{General Structure of CNNs}
\label{sec:General Structure of CNNs}

CNNs have become popular in recent years as they provide a computationally cheaper alternative to classical NNs.  They are commonly used in image recognition (e.g. \citealt{LeCun1999,Krizhevsky2012}), and have recently started being used in cosmology, for example, to find or analyze strong gravitational lenses in images (e.g. \citealt{Schaefer2017,Hezaveh2017}).

The first part of the CNN is the feature extractor (c.f. Fig. \ref{fig:CNN_scheme}).  Its purpose is to reduce the image, extracting patterns.\footnote{This is a form of data compression, analogous to the standard 21-cm analysis of performing an FFT and taking the PS, with the important difference that one does not {\it a priori} specify the convolution kernel but instead allows the network to learn it.}
The feature extraction is composed of successive convolution and pooling layers (c.f. Fig. \ref{fig:CNN_scheme}). 

A convolution layer takes an image as input and returns a series of concatenated convolutions analogous to a classical RGB (red-green-blue) image where the different convolutions represent the different colors.  The next layer perform a series of convolutions on the output of the previous one.  Those convolutions are in their turn concatenate in order to feed the next layer, etc.

Each convolution (also called a channel) is performed by an independent filter and is fed to an array of neurons (one per image pixel) each with their own activation function.  The number of filters, $\rm{ N_{f} }$, and their size, $\rm{ S_k }$, are the two hyper-parameters of the layer.  Each 2D filter is composed of $\rm{ S^2_k }$ trainable weights: these weights are adapted to the learning data.

A convolutional network is built by chaining convolution layers, each one working on the output of the previous one, after passing through the activation function, exactly as for a classical neural network.  The size of filters in one layer is fixed, therefore, one convolution layer can probe only one typical scale.  In order to access multi-scale information, the images are down-sampled by a pooling layer.  A pooling layer shrinks the image by keeping the maximal value in a kernel size $\rm{ S_{kMAX} }$, which is another hyper-parameter.  The function used in the pooling (here maximum) is also a hyper-parameter.

The sequence of convolutional and pooling layers defines the feature extraction part of the CNN. The resulting images are then flattened into a 1D array, by concatenating the rows from the final pooling layer. This 1D image then serves as an input into a \quotes{classical} neural network, used for parameter inference. This NN operates in so-called regression mode (i.e. used to predict continuous values of parameters, as opposed to categorizing).  Each layer of this NN is fully connected, with the final layer consisting of the four neurons corresponding to our astrophysical parameters. These four neurons are the only ones without activation functions.

\subsection{Training a neural network}
\label{sec:Training a neural network}

The training (or learning) of a network is the adaptation of all the weights as a function of how good the response of the network is for a given input.  The algorithm we describe here is general for NNs (CNNs do not require specific algorithms for training). 

Training a NN is an iterative process of guessing, computing the corresponding error and updating the network until the error becomes arbitrarily small. The guessing part is called \quotes{forward propagation}: running the network on the training sample. Then the error can be computed between the predicted ($\rm{y_{pred}}$) and the true values ($\rm{y_{true}}$) of the parameters. This error is quantified by a loss function. Finally, the updating of the weights is done by back-propagating in the network the loss information (the loss information goes from the parameters end to the input layer). Each iteration over the whole training sample is called an epoch, and the process of training to update the weights requires several epochs. In our study, the network needs $\sim$90 epochs to converge (see Fig. \ref{fig:CNN_loss} and associated discussion). 

There are many loss functions to choose from in the literature.  Here we use the mean square error defined by: $\rm{ MSE=\sum (y_{pred}-y_{true})^2 / N_y }$ where $\rm{N_y}$ is the data sample size and the summation occurs on the whole sample.  During training, the neural network tries to minimize this loss by changing the weights. The weights are updated by gradient descent: computing the derivative of the loss with respect to the weights, and following the gradient toward the minimum.

The updating of the weights is controlled by the \quotes{learning rate}. It controls how much the weights step from their current positions toward the minimum.  If the learning rate is too small, the convergence will be slow. If it is too large, the learning might never converge, as the weights keep stepping over the minimum.  Algorithms have been developed to optimize this process.  Here we use the RMSprop optimizer \citep{Hinton2012} in combination with a Batch Normalization layer \citep{Ioffe2015}. Additionally, we use an adaptive learning rate i.e. when the training loss does not evolve for a defined number of epochs (called 'patience'), the learning rate is decreased by a certain factor.  The patience and factor are additional hyper-parameters.

The learning phase can be further optimized by splitting the training sample into groups, called \quotes{mini-batches}. This allows the network to update its weights after only computing the loss from a mini-batch of training data (e.g. \citealt{Bengio2012, Masters2018}). The size of this mini-batch, $\rm{N_{batch}}$ is another hyper-parameter.  

\subsection{Tuning hyper-parameters and avoiding over-fitting}
\label{sec:Tuning hyper-parameters and avoiding over-fitting}

As shown in the previous sections, the building and training of a CNN include choosing a number of hyper-parameters.  For some of these, there can be physically-motivated choices.  For example, to reduce the required number of pooling layers, one can choose a convolutional filter size which is large enough to pick up the characteristic scales of the ionized and heated regions of the 21-cm signal.

However, most of the choices made in setting the CNN do not have obvious {\it a priori} motivations, and have to be \quotes{tuned} (e.g. \citealt{Bengio2012}). Tunning a NN is an iterative process, as for each new hyper-parameter the network can be re-trained and its recovery ability re-evaluated. As a result, tuning can be extremely computationally expensive.

In addition to the computational cost, one must be careful to avoid \quotes{over-fitting}. In choosing hyper-parameters which optimize performance on the training set, the network can become very specialized to the training data. If this over-fitting happens, the network will not be able to generalize the learned features of data outside of the training set.

\begin{figure}
\includegraphics[width=\columnwidth]{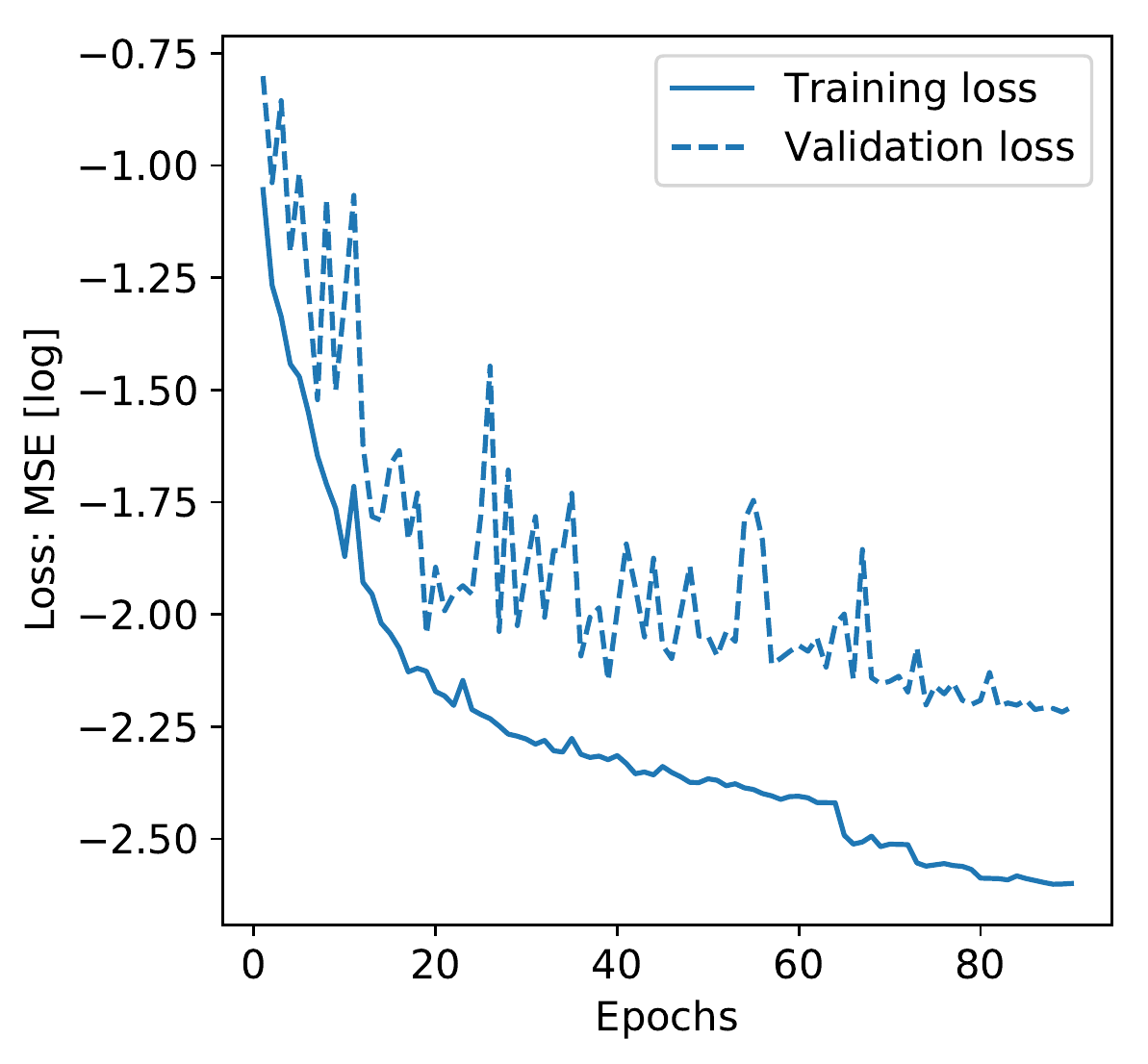}
\caption{ Loss of the CNN at each epoch during the training phase. The training loss (continuous line) is directly computed from the training sample. The validation loss (dashed line) is estimated at the end of each epoch with the validation sample (which is independent of the training and testing samples). }
\label{fig:CNN_loss}
\end{figure}

A common solution to avoid over-fitting is to set aside some fraction of the training data, and not use it during the learning phase when evaluating the hyper-parameters. This is called the \quotes{validation set} and is independent of the \quotes{testing set}, which is set aside and only used at the end in order to quantify the final performance of the CNN (see \S \ref{sec:Results: Parameter inference with a CNN}). In our case, the validation set consists of 1000 randomly-selected LC. The performance of the CNN on the validation set serves as a figure-of-merit for each chosen set of hyper-parameters.

In addition to using a validation sample to avoid over-fitting, we also turn off a random selection of 20\% of the neurons during training. This technique is called \quotes{dropout} \citep{Srivastava2014}, and it prevents the NN from relying too much on a fixed subset of information, preferring weights which include information from more neurons (so-called \quotes{regularization}).

We demonstrate that our CNN does not over-fit the training set in Fig. \ref{fig:CNN_loss}.  This figure presents the loss during learning, as a function of learning epoch, for our final CNN set-up discussed in the following section.
The solid line corresponds to the \quotes{training loss}, i.e. computed from the training sample and used in the back-propagation step of the learning. The dashed line corresponds to the \quotes{validation loss}, computed from the validation sample, but not used in the learning. As the network adapts, the losses decrease and at some point start to asymptote.  In our case, the validation loss asymptotes after $\sim$80 epochs. If the network were over-fitting, the training loss would continue to decrease, while the validation loss would start to {\it increase}: the network would start \quotes{over-specializing} on the training set.  This does not happen for our CNN; instead, the two losses show the same general trends.

\subsection{The final CNN set-up}
\label{sec:The final CNN set-up}

\renewcommand{\tabularxcolumn}[1]{>{\centering\arraybackslash}m{#1}}
\begin{table*}
\begin{center}
\begin{tabularx}{\textwidth}{XXXX}
\hline
(1) & (2) & (3) & (4) \\
Order & Layer / step name & Data dimension & Number of weights \\
\hline
\hline
1 & Input          & 100 $\times$ 2200            & .        \\ 
2 & 2D Convolution & 8  $\times$ 91 $\times$ 2191 & 808     \\
3 & Max Pooling    & 8  $\times$ 45 $\times$ 1095 & 0        \\
4 & 2D Convolution & 16 $\times$ 36 $\times$ 1086 & 12,816    \\
5 & Max Pooling    & 16 $\times$ 18 $\times$ 543  & 0        \\
6 & Flattening     & 156,384                       & 0        \\
7 & Dropout        & (20\% - only in learning phase) & 0        \\ 
8 & Fully connected      & 64           & 10,008,640 \\
9 & Batch Normalization  & .            & .        \\
10 & Fully connected      & 16           & 1040     \\
11 & Fully connected      & 8            & 136      \\
12 & Fully connected  Out & 4            & 36       \\
\hline
Total number of unknowns & . & 10,023,604 \\
\hline
Size convolution filter  & (10$\times$10) & \\
Size MaxPool filter      & (2$\times$2)   & \\
Activation function      & ReLU    & \\
Optimizer                & RMSprop & \\
Loss                     & MSE     & \\ 
Number of epochs         & 100     & \\
Batch size               & 200     & \\
Training set             & 8000$\times$10 &    \\
Validation set           & 1000    & \\
Testing set              & 1000    & \\
Learning rate patience   & 5 epochs& \\
Learning rate factor     & 0.5     & \\

\hline
\end{tabularx}
\end{center}
\caption{ Architecture and hyper-parameters of the Convolutional Neural Network used in this study. }
\label{tab:CNN}
\end{table*}

The final set-up of the CNN used in this study is presented in Tab. \ref{tab:CNN}.  The CNN is composed of two layers of convolution plus pooling.  The first convolution layer contains 8 filters and the second contains 16, with both using 10 x 10 filters.  Each pooling layer down-samples by a factor of 2.  The fully-connected part is composed of four layers with 64, 16, 8 and 4 neurons, respectively.  Our dropout fraction is 20 percent, applied on the layer that contains most of the weights: the one between the flattening layer and the first fully-connected layer. Our training / validation / test sets consists of (8,000$\times$10)\footnote{As mentioned previously, the training set contains 8,000 LC and we take 10 image slices per LC, resulting in 8000 $\times$ 10 images using in the training.   For validation and testing we only take one slice per LC as these are only used to check the performance of the inference and the size of the sets is already sufficient to obtain good statistics.} / 1,000 / 1,000 light-cone slices.  The CNN has been built using the python API for neural networks KERAS\footnote{https://github.com/keras-team/keras} \citep{Chollet2015}.

\section{Results: Parameter inference with a CNN}
\label{sec:Results: Parameter inference with a CNN}

Using the 1,000 test LCs, we now quantify the performance of the CNN described in the previous section.

\subsection{Coefficient of determination}
\label{sec:Coefficient of determination}

\renewcommand{\tabularxcolumn}[1]{>{\centering\arraybackslash}m{#1}}
\begin{table}
\centering
\begin{tabularx}{1\columnwidth}{XXX}
\hline
(1) & (2) & (3) \\
Parameter &  $\rm{R_2}$ score & $\rm{R_2}$ score  \\
 & (10 images / LC) & (1 image / LC) \\
\hline
\hline
\MTvir & 0.997 & 0.984 \\
\MLX   & 0.987 & 0.981 \\
\Mzeta & 0.955 & 0.851 \\
\MEo   & 0.728 & 0.531 \\
\hline
\end{tabularx}
\caption{ Coefficient of determination ($\rm{R_2}$) for our four astrophysical parameters, using the testing sample.  Columns (2) and (3) present results for two different training sets, using 10 and 1 slices per light-cone, respectively.  }
\label{tab:score_CNN}
\end{table}

We first evaluate the coefficient of determination for each of the four astrophysical parameters, $\rm{ R^2 }$, defined by: 

\begin{equation}
    \rm{ R^2 = \frac{ \sum (y_{pred} - \overline{y}_{true} )^2 }{ \sum (y_{true} - \overline{y}_{true} )^2 } = 1 - \frac{ \sum (y_{pred} - y_{true} )^2 }{ \sum (y_{true} - \overline{y}_{true} )^2 } },
\end{equation}

where $\rm{ \overline{y}_{true} }$ is the average of the true parameter and the summation is performed over the entire test set.  The $\rm{ R^2 }$ range can be between 0 and 1, where 1 indicates a perfect inference of the parameters. 

The resulting values of $\rm{ R^2 }$ are presented in Tab. \ref{tab:score_CNN} column (2). The parameters \MTvir and \MLX are almost perfectly inferred by the network with $\rm{R^2=0.99}$.  The parameter \Mzeta results in a slightly lower score, $\rm{R^2=0.95}$, which is still good.  Finally, the last parameter \MEo has a comparably low score, $\rm{R^2=0.73}$.  We shall explore the reason for this in the following section.

We additionally present in Tab. \ref{tab:score_CNN} column (3) the scores resulting after learning is performed on a training set which is reduced in size by a factor of 10: consisting of only one slice per LC.  As expected, a smaller training set worsens the recovery, especially for the parameters \Mzeta and \MEo. This indicates that we could expect even better performance if we were to use the entire LC in the training (corresponding to 400 of these 2D images), instead of throwing away 39/40 of the data as we are doing in our fiducial CNN. We plan on investigating this further in follow-up work, taking advantage of GPU-accelerated ML packages.

\subsection{Predicted vs True distributions}
\label{sec:Inferred parameters}

\begin{figure*}
\includegraphics[width=\columnwidth]{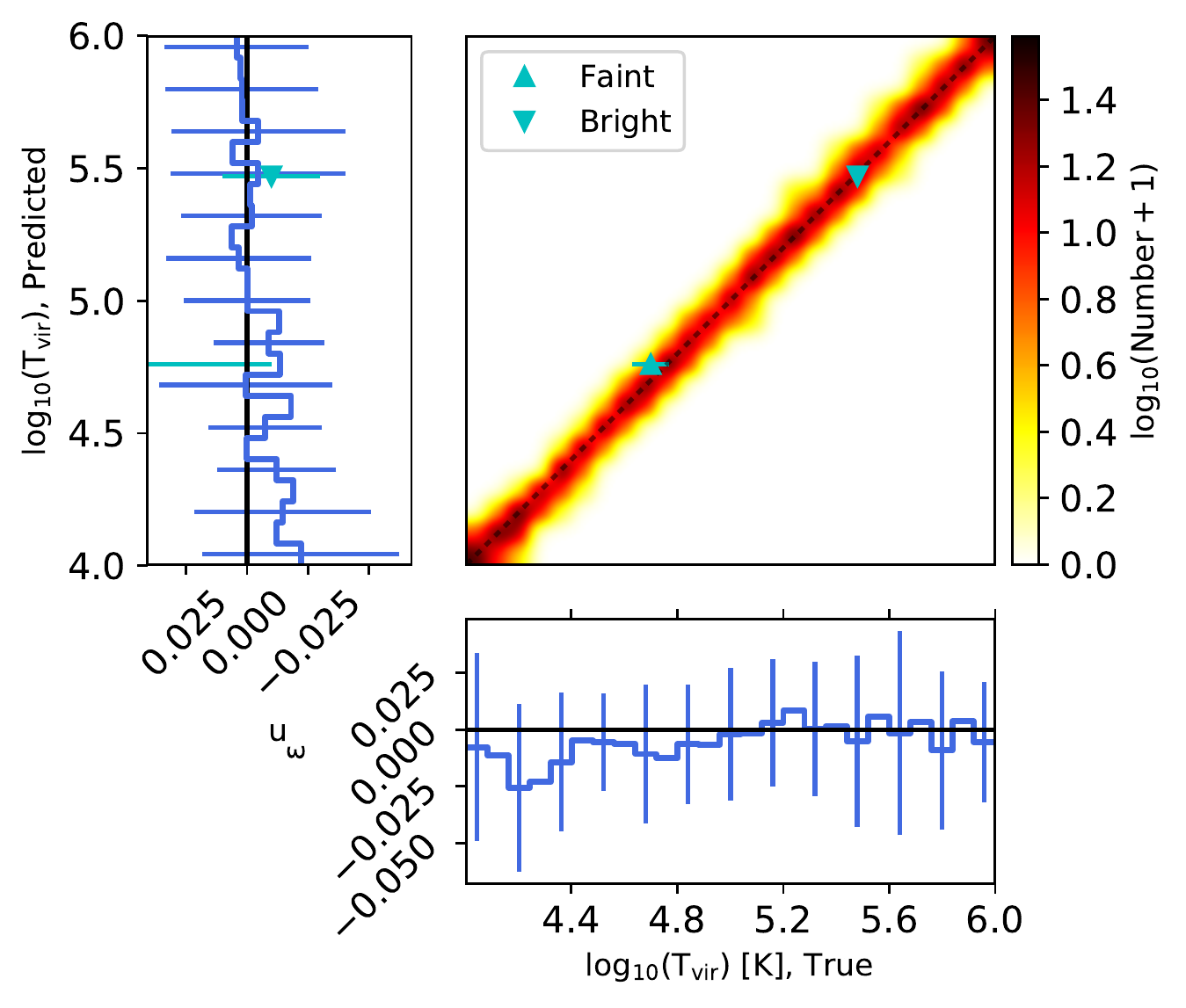}
\includegraphics[width=\columnwidth]{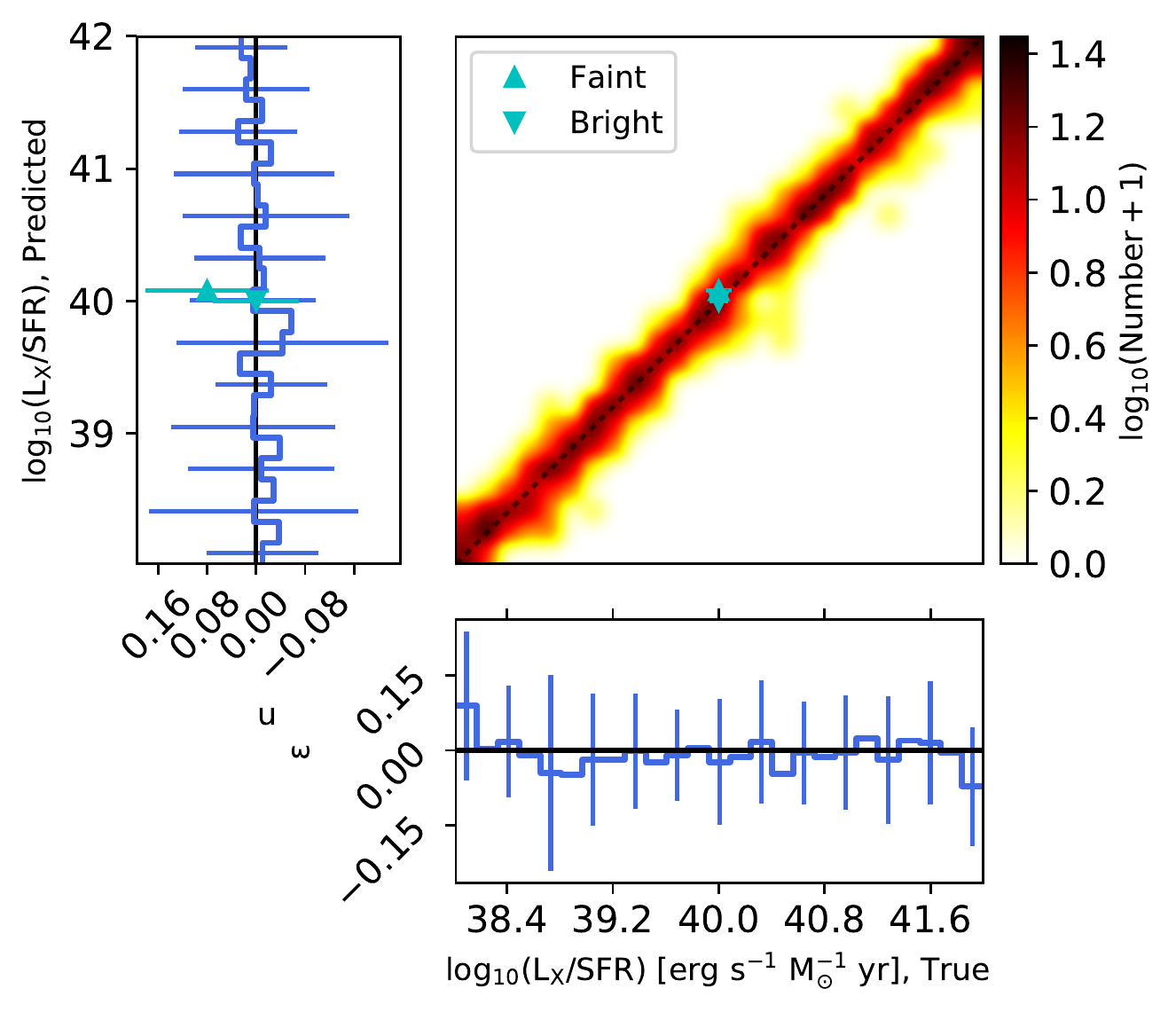} \\
\includegraphics[width=\columnwidth]{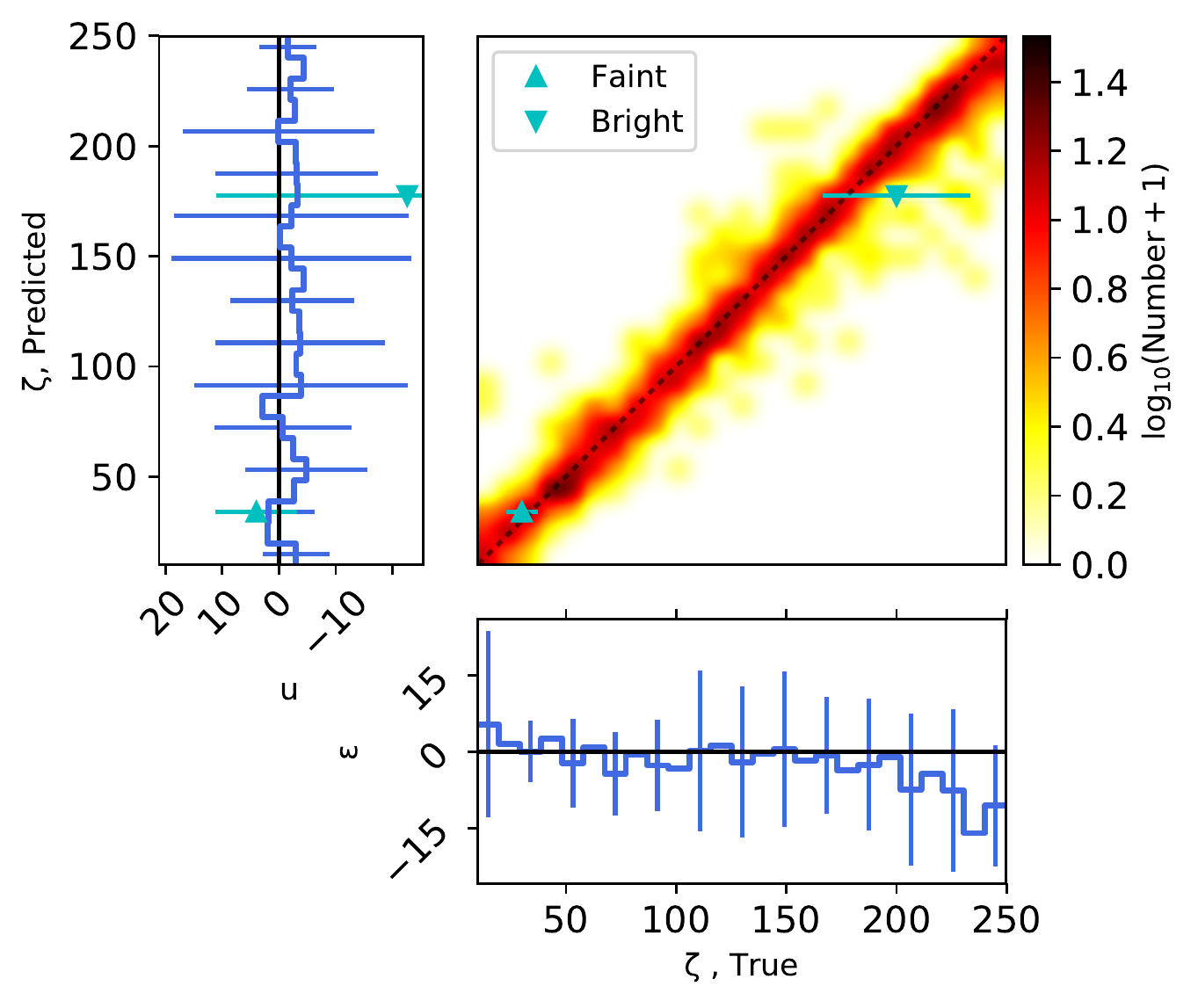} 
\includegraphics[width=\columnwidth]{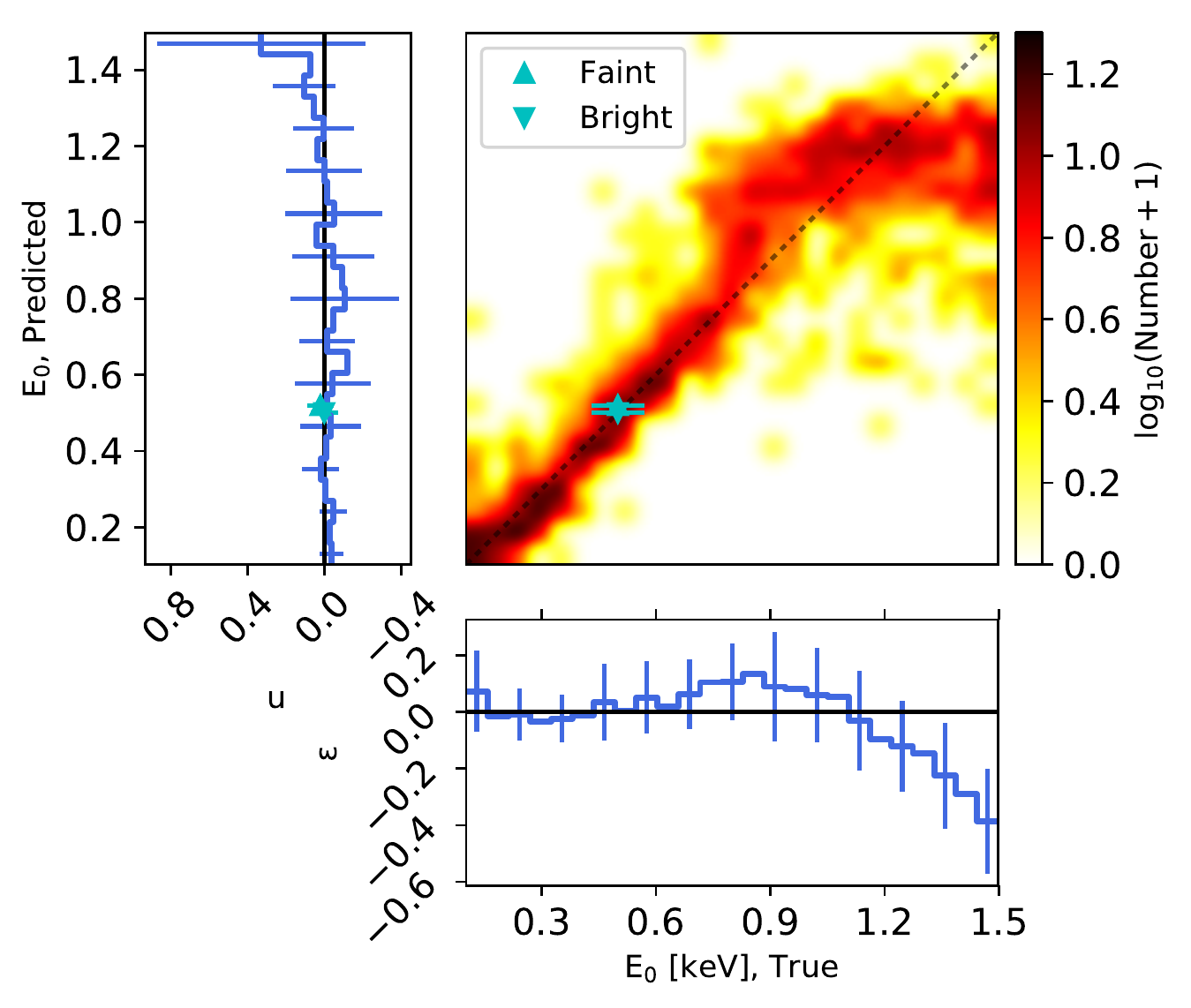}
\caption{ CNN parameter inference using the testing sample which contains 1000 LC. The central image in each panel contains the distributions of true vs predicted parameter values, $\rm {p (y_{\rm pred}, y_{\rm true})}$, with the side color bar indicating the log number count in each bin. The closer the points are to the diagonal, the better is the performance. The bottom and left histograms present the mean and the standard deviation (taking a Gaussian fit) of the residual, $\rm{y_{pred} - y_{true}}$. The bottom histogram is the learning error, $\rm{ p( y_{pred} - y_{true} | y_{true}) }$, while the side histogram is the recovery uncertainty, $\rm{ p( y_{pred} -y_{true} | y_{pred}) }$. The two triangle points mark the position and the marginalized error for MCMC analysis of \citetalias{Greig2018} using \code{21CMMC}.  }
\label{fig:2D_results}
\end{figure*}

The $\rm{ R^2 }$ score from the previous section gives an estimate of the CNN's performance averaged over the whole parameter range.  Here we go into more detail by presenting the distributions of predicted vs true parameter values in Fig. \ref{fig:2D_results}.  This figure is the main result of this work.

For each parameter, the central panel shows the density of points (see adjacent color bar) in the corresponding 2D space. The bottom and left side histograms present the mean and standard deviation of the residual $\rm{ r \equiv y_{pred} -y_{true} }$, along the vertical and horizontal directions, respectively.  Specifically, the bottom histogram shows the so-called network error, $\epsilon$, i.e. the distribution of residuals as a function of the true values, $\rm{ p( r | y_{true}) }$.  The left histogram shows the prediction uncertainty, $u$, i.e. the distribution of residuals  as a function of the prediction, $\rm{ p( r | y_{pred}) }$.  In practice, the left panel is the most relevant for quantifying the performance, since we will eventually have a single 21-cm LC observation of our single Universe with the \quotes{true} values unknown to us. Thus we are interested in what is the allowed range of true values, given the prediction we will have from the CNN.

We also denote with cyan points and error bars the recovered values and marginalized error from the MCMC analysis in \citetalias{Greig2018}. Those authors performed recovery using two mock observations: \quotes{Faint} and \quotes{Bright} galaxies, with correspondingly different values of $\zeta$ and $T_{\rm vir}$.  We caution that it is difficult to directly compare against these results since: (i) the marginalized uncertainty from \citetalias{Greig2018} does not directly translate to our recovery uncertainty; and (ii), unlike \citetalias{Greig2018}, we do not include instrument noise.  As such, quantitative comparison of the uncertainties only makes sense if they are not noise dominated, but are intrinsic to the theoretical model and parameter degeneracies.  As we shall see below, the uncertainties in both studies follow similar qualitative trends. As this work is only a proof-of-concept, we defer more detailed comparisons, including various telescope noise and point spread function (PSF) models for future work.

For the parameters \MTvir and \MLX (top row), the distributions are tightly centered along the diagonal over the entire range. There are no obvious biases and the prediction uncertainties shown in the left side panels are $ \rm{ \sigma_{T_{vir}} = 0.028 }$ and $\rm{ \sigma_{L_X} = 0.11 }$. These uncertainties are roughly constant over the entire parameter range and are comparable to those from \citetalias{Greig2018} for the two \quotes{true} values used in that work. \footnote{One of the benefits of using ML for parameter inference is that we can immediately evaluate the network's performance over the entire parameter range (c.f. the left side panels), without having to perform a separate MCMC for each mock observation. Obviously, this is only relevant before we have an actual observation, which will correspond to a single Universe.}

The recovery of \Mzeta (bottom left panel), is in general good though worse than the previous two parameters. Unlike for the previous two parameters, the recovery uncertainty is not constant over the whole parameter range. Consistent with the MCMC analysis of \citetalias{Greig2018},  high values of $\zeta$ (their Bright Galaxies model) have larger uncertainties than low values (their Faint Galaxies model). An explanation of this could be the following. \Mzeta controls the timing of reionization. Reionization driven by rare, bright galaxies must be rapid, requiring very high ionizing efficiencies, \Mzeta, in order to match the constraints from {\it Planck} and high-$z$ QSOs.  Having increasingly rapid reionization compresses the corresponding 21-cm signal into an increasingly narrow redshift range, decreasing and saturating how much information is available for inference.  As for the previous parameters, the size of the errors from the CNN is comparable to the MCMC analysis which uses the PS.

Finally, in the bottom right panel, we quantify the CNN's performance in recovering \MEo.  As expected from its comparably low coefficient of determination, $\rm{ R^2\sim 0.7 }$, the recovery is modest.  But despite this low score, we can see that the parameter is relatively well learned on the lower half of the parameter range ($\rm{ E_0 \in [0.1,0.8] }$ keV), while on the higher part it seems to saturate.  Note, that beyond 0.8 keV we no longer trust our reported errors because of the saturation. The conclusion here is that above $\sim$0.8 keV the results are not trustworthy.  This saturation is understandable.  High values of \MEo imply a harder X-ray spectrum.  Because the ionization cross-sections of hydrogen and helium are very strong functions of energy, $\sigma \propto E^{-3}$, harder X-rays interact increasingly less with the IGM and thus leave an increasingly smaller imprint in the light-cone.  Indeed beyond energies of $\sim$1.5 keV, the mean free path of X-rays surpasses the Hubble length and so we should not be able to distinguish any model with \MEo $\geq $ 1.5 keV.  It would be interesting to perform a parameter recovery using \code{21CMMC} with a mock observation having $\rm{ E_0 \sim 1.2\ keV }$ to confirm this trend. 

\section{Discussion and Conclusions}
\label{sec:Discussions and Conclusions}

The advent of next-generation 21-cm interferometers, like HERA and SKA, will allow us to create 3D images of the first billion years of our Universe.  We need sophisticated analysis tools in order to interpret these observations and understand the underlying astrophysics. Recent effort has been placed on astrophysical parameter recovery, using Bayesian approaches like MCMC. These require assuming a summary statistic when computing the likelihood, with the power spectrum being an obvious choice. Although the PS was shown to be a powerful discriminant of reionization and cosmic dawn astrophysical parameters, the 21-cm tomographic signal is highly non-Gaussian (unlike the CMB). Therefore there is additional information which is wasted if only the PS is used for parameter recovery.

In this proof-of-concept paper, we demonstrate that astrophysical parameters can be recovered directly from 21-cm images using deep learning with convolutional neural networks. CNNs are able to adapt to the training data, without requiring the user to {\it a priori} specify a summary statistic. However this flexibility comes at a cost of physical intuition (see the appendix for more details), and also makes having a fully Bayesian interpretation more difficult.  This statement comes from the fact that the CNN in inference mode (regression) only returns the best guess (analogous to a maximum likelihood), and not the probability of the result. The inferred marginalized errors of an observation can be estimated from Fig. \ref{fig:2D_results}, effectively marginalizing over a sample of "best-guesses" but this is not a posterior (as for a classical MCMC). The posterior information is encoded in all neurons' weights, but quantifying this is non-intuitive. \cite{Levasseur2017} provide a framework to estimate the final error made on parameter inference, taking into account the noise from the data and the error of the network itself.  It requires different kind of neural networks than the CNN used in this study, but it should be used for future networks that will be deployed on the observations.

We train our CNN using a database of 10,000 light-cones, of which 2,000 are reserved for validation and testing.  We demonstrate that our tuned network does not over-fit the training set.  Using 10 2D image slices per light-cone for training, we show that the CNN is able to recover popular parameters describing the first galaxies: (i) \MTvir, the minimum host halo mass capable of hosting efficient star formation; (ii) \Mzeta, their typical ionizing efficiencies; (iii) \MLX  their typical soft-band X-ray luminosity to star formation rate; and (iv) \MEo, the minimum X-ray energy capable of escaping into the IGM (governed by their mean HI column densities). 

For most of their allowed ranges, parameters \MTvir and \MLX are recovered with $< 1\%$  uncertainty, while  \Mzeta and \MEo are recovered with $\sim 10\%$ uncertainty.  Our results are roughly comparable to the accuracy obtained from Monte Carlo Markov Chain sampling of the PS with \code{21CMMC} for the two mock observations analyzed previously.  Although we caution that we do not yet include noise in this proof-of-concept study.  

Moreover, the computational cost of parameter inference using a CNN is distributed differently than using an MCMC code. Training the network requires (i) generating a large database of simulated LCs, and (ii) tuning hyper-parameters. In the case of this study, each of these steps took a couple of weeks of computation time on a dedicated server. However, once the database is in place and the network tuned, performing parameter inference on an \quotes{observation} is extremely fast, requiring only a few seconds per light-cone. In contrast, \code{21CMMC} requires no tuning; however, parameter inference for each new \quotes{observation} requires a comparable number ($\sim$10,000) of light-cone samples. This \quotes{up front} computational cost makes neural networks useful when one wishes to study parameter inference for a large number of mock observations before we have an actual observation in-hand. Indeed, with our CNN test sample we demonstrated that the recovered accuracy of \Mzeta and especially \MEo depend strongly on the \quotes{true} parameters. In particular, we show that the recovery of $E_0$ starts to worsen at energies $\gtrsim 0.8$ keV, as such hard X-ray photons do not interact strongly with the IGM.

It is difficult to {\it a priori} estimate how much these idealized conditions affect the results of our parameter recovery. However, we note that we only used a small fraction of the current information content of the LCs, due to the high computational costs associated with training and optimizing the CNN.  This reserve of data could offset the effective signal loss from noise and systematics.  Indeed, in a paper which appeared subsequent to the completion of this work, \cite{LaPlante2018} showed that an empirical parameter like the duration of reionization, can be recovered to 10\% precision in the presence of noise from 3D LC data.  In future work, we will investigate astrophysical parameters recovery in the presence of realistic noise and systematics.  

\section*{Acknowledgements}
We thank B. Semelin for its useful comments and discussions. This work was supported by the European Research Council (ERC) under the European Union's Horizon 2020 research and innovation programme (grant agreement No 638809 -- AIDA -- PI: Mesinger). The results presented here reflect the authors' views; the ERC is not responsible for their use.
Parts of this research were supported by the Australian Research Council Centre of Excellence for All Sky Astrophysics in 3 Dimensions (ASTRO 3D), through project number CE170100013.
AL acknowledges support for this work by NASA through Hubble Fellowship grant \#HST-HF2-51363.001-A awarded by the Space Telescope Science Institute, which is operated by the Association of Universities for Research in Astronomy, Inc., for NASA, under contract NAS5-26555.
We acknowledge support from INAF under PRIN SKA/CTA FORECaST.
We thank contributors to SciPy\footnote{http://www.scipy.org}, Matplotlib\footnote{http://www.matplotlib.sourceforge.net}, pyDOE\footnote{https://pythonhosted.org/pyDOE/}, and the Python programming language\footnote{http://www.python.org}.



\bibliographystyle{mnras}
\bibliography{Mendeley_LEoR}

\begin{thebibliography}{}
\makeatletter
\relax
\def\mn@urlcharsother{\let\do\@makeother \do\$\do\&\do\#\do\^\do\_\do\%\do\~}
\def\mn@doi{\begingroup\mn@urlcharsother \@ifnextchar [ {\mn@doi@}
  {\mn@doi@[]}}
\def\mn@doi@[#1]#2{\def\@tempa{#1}\ifx\@tempa\@empty \href
  {http://dx.doi.org/#2} {doi:#2}\else \href {http://dx.doi.org/#2} {#1}\fi
  \endgroup}
\def\mn@eprint#1#2{\mn@eprint@#1:#2::\@nil}
\def\mn@eprint@arXiv#1{\href {http://arxiv.org/abs/#1} {{\tt arXiv:#1}}}
\def\mn@eprint@dblp#1{\href {http://dblp.uni-trier.de/rec/bibtex/#1.xml}
  {dblp:#1}}
\def\mn@eprint@#1:#2:#3:#4\@nil{\def\@tempa {#1}\def\@tempb {#2}\def\@tempc
  {#3}\ifx \@tempc \@empty \let \@tempc \@tempb \let \@tempb \@tempa \fi \ifx
  \@tempb \@empty \def\@tempb {arXiv}\fi \@ifundefined
  {mn@eprint@\@tempb}{\@tempb:\@tempc}{\expandafter \expandafter \csname
  mn@eprint@\@tempb\endcsname \expandafter{\@tempc}}}

\bibitem[\protect\citeauthoryear{Ade et~al.,}{Ade
  et~al.}{2016}]{PlanckCollaboration2015}
Ade P. A.~R.,  et~al., 2016, \mn@doi [Astronomy {\&} Astrophysics]
  {10.1051/0004-6361/201525830}, 594, A13

\bibitem[\protect\citeauthoryear{Barkana \& Loeb}{Barkana \&
  Loeb}{2008}]{Barkana2008}
Barkana R.,  Loeb A.,  2008, \mn@doi [Monthly Notices of the Royal Astronomical
  Society] {10.1111/j.1365-2966.2007.12729.x}, 384, 1069

\bibitem[\protect\citeauthoryear{Bengio}{Bengio}{2012}]{Bengio2012}
Bengio Y.,  2012, \mn@doi [Lecture Notes in Computer Science (including
  subseries Lecture Notes in Artificial Intelligence and Lecture Notes in
  Bioinformatics)] {10.1007/978-3-642-35289-8-26}, 7700 LECTU, 437

\bibitem[\protect\citeauthoryear{Bharadwaj \& Pandey}{Bharadwaj \&
  Pandey}{2005}]{Bharadwaj2005}
Bharadwaj S.,  Pandey S.~K.,  2005, \mn@doi [Monthly Notices of the Royal
  Astronomical Society] {10.1111/j.1365-2966.2005.08836.x}, 358, 968

\bibitem[\protect\citeauthoryear{Chollet}{Chollet}{2015}]{Chollet2015}
Chollet F.,  2015, {Keras}, \url {https://github.com/keras-team/keras}

\bibitem[\protect\citeauthoryear{Das, Mesinger, Pallottini, Ferrara  \&
  Wise}{Das et~al.}{2017}]{Das2017}
Das A.,  Mesinger A.,  Pallottini A.,  Ferrara A.,   Wise J.~H.,  2017, \mn@doi
  [Monthly Notices of the Royal Astronomical Society] {10.1093/mnras/stx943},
  469, 1166

\bibitem[\protect\citeauthoryear{Deboer et~al.,}{Deboer
  et~al.}{2017}]{Deboer2017}
Deboer D.~R.,  et~al., 2017, \mn@doi [Publications of the Astronomical Society
  of the Pacific] {10.1088/1538-3873/129/974/045001}, 129

\bibitem[\protect\citeauthoryear{Fialkov, Cohen, Barkana  \& Silk}{Fialkov
  et~al.}{2017}]{Fialkov2017}
Fialkov A.,  Cohen A.,  Barkana R.,   Silk J.,  2017, \mn@doi [Monthly Notices
  of the Royal Astronomical Society] {10.1093/mnras/stw2540}, 464, 3498

\bibitem[\protect\citeauthoryear{Field}{Field}{1958}]{Field1958}
Field G.,  1958, \mn@doi [Proceedings of the IRE] {10.1109/JRPROC.1958.286741},
  46, 240

\bibitem[\protect\citeauthoryear{Furlanetto, Zaldarriaga  \&
  Hernquist}{Furlanetto et~al.}{2004}]{Furlanetto2004}
Furlanetto S.~R.,  Zaldarriaga M.,   Hernquist L.,  2004, \mn@doi [The
  Astrophysical Journal] {10.1086/423028}, 613, 16

\bibitem[\protect\citeauthoryear{Greig \& Mesinger}{Greig \&
  Mesinger}{2015}]{Greig2015}
Greig B.,  Mesinger A.,  2015, \mn@doi [Monthly Notices of the Royal
  Astronomical Society] {10.1093/mnras/stv571}, 449, 4246

\bibitem[\protect\citeauthoryear{Greig \& Mesinger}{Greig \&
  Mesinger}{2017}]{Greig2017}
Greig B.,  Mesinger A.,  2017, \mn@doi [Monthly Notices of the Royal
  Astronomical Society] {10.1093/mnras/stx2118}, 472, 2651

\bibitem[\protect\citeauthoryear{Greig \& Mesinger}{Greig \&
  Mesinger}{2018}]{Greig2018}
Greig B.,  Mesinger A.,  2018, Mon. Not. R. Astron. Soc, 000, 0

\bibitem[\protect\citeauthoryear{Gupta, Matilla, Hsu  \& Haiman}{Gupta
  et~al.}{2018}]{Gupta2018}
Gupta A.,  Matilla J. M.~Z.,  Hsu D.,   Haiman Z.,  2018, preprint

\bibitem[\protect\citeauthoryear{Haarlem, Wise, Gunst, Heald, Mckean, Hessels
  \& Bruyn}{Haarlem et~al.}{2013}]{Haarlem2013}
Haarlem M. P.~V.,  Wise M.~W.,  Gunst A.~W.,  Heald G.,  Mckean J.~P.,  Hessels
  J. W.~T.,   Bruyn A. G.~D.,  2013, \mn@doi [Astronomy and Astrophysics]
  {10.1051/0004-6361/201220873}, 556, A2

\bibitem[\protect\citeauthoryear{Hezaveh, Levasseur  \& Marshall}{Hezaveh
  et~al.}{2017}]{Hezaveh2017}
Hezaveh Y.~D.,  Levasseur L.~P.,   Marshall P.~J.,  2017, \mn@doi [Nature]
  {10.1038/nature23463}, 548, 555

\bibitem[\protect\citeauthoryear{Hinton, Srivastava  \& Swersky}{Hinton
  et~al.}{2012}]{Hinton2012}
Hinton G.~E.,  Srivastava N.,   Swersky K.,  2012, COURSERA: Neural Networks
  for Machine Learning, p.~31

\bibitem[\protect\citeauthoryear{Iliev, Mellema, Shapiro, Pen, Mao, Koda  \&
  Ahn}{Iliev et~al.}{2011}]{Iliev2012}
Iliev I.~T.,  Mellema G.,  Shapiro P.~R.,  Pen U.-L.,  Mao Y.,  Koda J.,   Ahn
  K.,  2011, \mn@doi [Monthly Notices of the Royal Astronomical Society]
  {10.1111/j.1365-2966.2012.21032.x}, 423, 2222

\bibitem[\protect\citeauthoryear{Iliev, Mellema, Ahn, Shapiro, Mao  \&
  Pen}{Iliev et~al.}{2014}]{Iliev2014}
Iliev I.~T.,  Mellema G.,  Ahn K.,  Shapiro P.~R.,  Mao Y.,   Pen U.-L.,  2014,
  \mn@doi [Monthly Notices of the Royal Astronomical Society]
  {10.1093/mnras/stt2497}, 439, 725

\bibitem[\protect\citeauthoryear{Ioffe \& Szegedy}{Ioffe \&
  Szegedy}{2015}]{Ioffe2015}
Ioffe S.,  Szegedy C.,  2015, preprint

\bibitem[\protect\citeauthoryear{Kamdar, Turk  \& Brunner}{Kamdar
  et~al.}{2016a}]{Kamdar2016}
Kamdar H.~M.,  Turk M.~J.,   Brunner R.~J.,  2016a, \mn@doi [Monthly Notices of
  the Royal Astronomical Society] {10.1093/mnras/stv2310}, 455, 642

\bibitem[\protect\citeauthoryear{Kamdar, Turk  \& Brunner}{Kamdar
  et~al.}{2016b}]{Kamdar2016a}
Kamdar H.~M.,  Turk M.~J.,   Brunner R.~J.,  2016b, \mn@doi [Monthly Notices of
  the Royal Astronomical Society] {10.1093/mnras/stv2981}, 457, 1162

\bibitem[\protect\citeauthoryear{Kern, Liu, Parsons, Mesinger  \& Greig}{Kern
  et~al.}{2017}]{Kern2017}
Kern N.~S.,  Liu A.,  Parsons A.~R.,  Mesinger A.,   Greig B.,  2017, \mn@doi
  [The Astrophysical Journal] {10.3847/1538-4357/aa8bb4}, 848, 23

\bibitem[\protect\citeauthoryear{Koopmans et~al.,}{Koopmans
  et~al.}{2015}]{Koopmans2015}
Koopmans L.,  et~al., 2015, \mn@doi [Proceedings of Advancing Astrophysics with
  the Square Kilometre Array (AASKA14). 9 -13 June] {2015aska.confE...1K},
  9-13-June-, 1

\bibitem[\protect\citeauthoryear{Krizhevsky, Sutskever  \& Hinton}{Krizhevsky
  et~al.}{2012}]{Krizhevsky2012}
Krizhevsky A.,  Sutskever I.,   Hinton G.~E.,  2012, \mn@doi [Advances In
  Neural Information Processing Systems]
  {http://dx.doi.org/10.1016/j.protcy.2014.09.007}, pp~1--9

\bibitem[\protect\citeauthoryear{La~Plante \& Ntampaka}{La~Plante \&
  Ntampaka}{2018}]{LaPlante2018}
La~Plante P.,  Ntampaka M.,  2018, preprint, 1810.08211

\bibitem[\protect\citeauthoryear{LeCun, Haffner, Bottou  \& Bengio}{LeCun
  et~al.}{1999}]{LeCun1999}
LeCun Y.,  Haffner P.,  Bottou L.,   Bengio Y.,  1999, Springer, Berlin,
  Heidelberg, pp 319--345, \mn@doi{10.1007/3-540-46805-6{\_}19}, \url
  {http://link.springer.com/10.1007/3-540-46805-6_19}

\bibitem[\protect\citeauthoryear{Levasseur, Hezaveh  \& Wechsler}{Levasseur
  et~al.}{2017}]{Levasseur2017}
Levasseur L.~P.,  Hezaveh Y.~D.,   Wechsler R.~H.,  2017, \mn@doi
  [Astrophysical Journal Letters] {10.3847/2041-8213/aa9704}, 850

\bibitem[\protect\citeauthoryear{Majumdar, Pritchard, Mondal, Watkinson,
  Bharadwaj  \& Mellema}{Majumdar et~al.}{2018}]{Majumdar2018}
Majumdar S.,  Pritchard J.~R.,  Mondal R.,  Watkinson C.~A.,  Bharadwaj S.,
  Mellema G.,  2018, \mn@doi [Monthly Notices of the Royal Astronomical
  Society] {10.1093/mnras/sty535}, 476, 4007

\bibitem[\protect\citeauthoryear{Masters \& Luschi}{Masters \&
  Luschi}{2018}]{Masters2018}
Masters D.,  Luschi C.,  2018, preprint, 1804.07612

\bibitem[\protect\citeauthoryear{McQuinn}{McQuinn}{2012}]{McQuinn2012}
McQuinn M.,  2012, \mn@doi [Monthly Notices of the Royal Astronomical Society]
  {10.1111/j.1365-2966.2012.21792.x}, 426, 1349

\bibitem[\protect\citeauthoryear{McQuinn, Lidz, Zahn, Dutta, Hernquist  \&
  Zaldarriaga}{McQuinn et~al.}{2007}]{McQuinn2007}
McQuinn M.,  Lidz A.,  Zahn O.,  Dutta S.,  Hernquist L.,   Zaldarriaga M.,
  2007, \mn@doi [Monthly Notices of the Royal Astronomical Society]
  {10.1111/j.1365-2966.2007.11489.x}, 377, 1043

\bibitem[\protect\citeauthoryear{Mellema et~al.,}{Mellema
  et~al.}{2013}]{Mellema}
Mellema G.,  et~al., 2013, \mn@doi [Experimental Astronomy]
  {10.1007/s10686-013-9334-5}, 36, 235

\bibitem[\protect\citeauthoryear{Mellema, Koopmans, Shukla, Datta, Mesinger,
  Majumdar  \& Group}{Mellema et~al.}{2015}]{Mellema2014}
Mellema G.,  Koopmans L.,  Shukla H.,  Datta K.~K.,  Mesinger A.,  Majumdar S.,
    Group o. b. o. t. C. S.~W.,  2015, Advancing Astrophysics with the Square
  Kilometre Array (AASKA14)

\bibitem[\protect\citeauthoryear{Mesinger \& Furlanetto}{Mesinger \&
  Furlanetto}{2007}]{Mesinger2007}
Mesinger A.,  Furlanetto S.,  2007, \mn@doi [The Astrophysical Journal]
  {10.1086/521806}, 669, 663

\bibitem[\protect\citeauthoryear{Mesinger, Furlanetto  \& Cen}{Mesinger
  et~al.}{2011}]{Mesinger2011}
Mesinger A.,  Furlanetto S.,   Cen R.,  2011, \mn@doi [Monthly Notices of the
  Royal Astronomical Society] {10.1111/j.1365-2966.2010.17731.x}, 411, 955

\bibitem[\protect\citeauthoryear{Nair \& Hinton}{Nair \&
  Hinton}{2010}]{Nair2010}
Nair V.,  Hinton G.~E.,  2010, \mn@doi [Proceedings of the 27th International
  Conference on Machine Learning] {10.1.1.165.6419}, pp 807--814

\bibitem[\protect\citeauthoryear{Pacucci, Mesinger, Mineo  \& Ferrara}{Pacucci
  et~al.}{2014}]{Pacucci2014}
Pacucci F.,  Mesinger A.,  Mineo S.,   Ferrara A.,  2014, \mn@doi [Monthly
  Notices of the Royal Astronomical Society] {10.1093/mnras/stu1240}, 443, 678

\bibitem[\protect\citeauthoryear{Parks, Prochaska, Dong  \& Cai}{Parks
  et~al.}{2018}]{Parks2017}
Parks D.,  Prochaska J.~X.,  Dong S.,   Cai Z.,  2018, \mn@doi [Monthly Notices
  of the Royal Astronomical Society] {10.1093/mnras/sty196}, 476, 1151

\bibitem[\protect\citeauthoryear{Parsons et~al.,}{Parsons
  et~al.}{2010}]{Parsons2010}
Parsons A.~R.,  et~al., 2010, \mn@doi [Astronomical Journal]
  {10.1088/0004-6256/139/4/1468}, 139, 1468

\bibitem[\protect\citeauthoryear{Rodriguez, Kacprzak, Lucchi, Amara, Sgier,
  Fluri, Hofmann  \& R{\'{e}}fr{\'{e}}gier}{Rodriguez
  et~al.}{2018}]{Rodriguez2018}
Rodriguez A.~C.,  Kacprzak T.,  Lucchi A.,  Amara A.,  Sgier R.,  Fluri J.,
  Hofmann T.,   R{\'{e}}fr{\'{e}}gier A.,  2018, preprint

\bibitem[\protect\citeauthoryear{Schaefer, Geiger, Kuntzer  \& Kneib}{Schaefer
  et~al.}{2017}]{Schaefer2017}
Schaefer C.,  Geiger M.,  Kuntzer T.,   Kneib J.-P.,  2017, \mn@doi [Astronomy
  {\&} Astrophysics] {10.1051/0004-6361/201731201}, 611

\bibitem[\protect\citeauthoryear{Schmit \& Pritchard}{Schmit \&
  Pritchard}{2018}]{Schmit2018}
Schmit C.~J.,  Pritchard J.~R.,  2018, \mn@doi [Monthly Notices of the Royal
  Astronomical Society] {10.1093/mnras/stx3292}, 475, 1213

\bibitem[\protect\citeauthoryear{Shimabukuro \& Semelin}{Shimabukuro \&
  Semelin}{2017}]{Shimabukuro2017}
Shimabukuro H.,  Semelin B.,  2017, \mn@doi [MNRAS Preprint]
  {10.1093/mnras/stx734}, 000, 1

\bibitem[\protect\citeauthoryear{Sobacchi \& Mesinger}{Sobacchi \&
  Mesinger}{2015}]{Sobacchi2015}
Sobacchi E.,  Mesinger A.,  2015, \mn@doi [Monthly Notices of the Royal
  Astronomical Society] {10.1093/mnras/stv1751}, 453, 1843

\bibitem[\protect\citeauthoryear{Srivastava, Hinton, Krizhevsky, Sutskever  \&
  Salakhutdinov}{Srivastava et~al.}{2014}]{Srivastava2014}
Srivastava N.,  Hinton G.,  Krizhevsky A.,  Sutskever I.,   Salakhutdinov R.,
  2014, \mn@doi [Journal of Machine Learning Research] {10.1214/12-AOS1000},
  15, 1929

\bibitem[\protect\citeauthoryear{Tingay et~al.,}{Tingay
  et~al.}{2013}]{Tingay2013}
Tingay S.~J.,  et~al., 2013, \mn@doi [Publications of the Astronomical Society
  of Australia] {10.1017/pasa.2012.007}, 30

\bibitem[\protect\citeauthoryear{Ucci, Ferrara, Gallerani  \& Pallottini}{Ucci
  et~al.}{2016}]{Ucci2017}
Ucci G.,  Ferrara A.,  Gallerani S.,   Pallottini A.,  2016, \mn@doi [Monthly
  Notices of the Royal Astronomical Society] {10.1093/mnras/stw2836}, 465, 1144

\bibitem[\protect\citeauthoryear{Ucci, Ferrara, Pallottini  \& Gallerani}{Ucci
  et~al.}{2018}]{Ucci2018}
Ucci G.,  Ferrara A.,  Pallottini A.,   Gallerani S.,  2018, \mn@doi [Monthly
  Notices of the Royal Astronomical Society] {10.1093/mnras/sty804}, 477, 1484

\bibitem[\protect\citeauthoryear{Watkinson \& Pritchard}{Watkinson \&
  Pritchard}{2015}]{Watkinson2018}
Watkinson C.~A.,  Pritchard J.~R.,  2015, \mn@doi [Monthly Notices of the Royal
  Astronomical Society] {10.1093/mnras/stv2010}, 454, 1416

\bibitem[\protect\citeauthoryear{Wouthuysen}{Wouthuysen}{1952}]{Wouthuysen1952}
Wouthuysen S.~A.,  1952, \mn@doi [The Astronomical Journal] {10.1086/106661},
  57, 31

\bibitem[\protect\citeauthoryear{Yatawatta et~al.,}{Yatawatta
  et~al.}{2013}]{Yatawatta2013}
Yatawatta S.,  et~al., 2013, \mn@doi [Astronomy and Astrophysics]
  {10.1051/0004-6361/201220874}, 550, 136

\bibitem[\protect\citeauthoryear{Zahn, Lidz, McQuinn, Dutta, Hernquist,
  Zaldarriaga  \& Furlanetto}{Zahn et~al.}{2007}]{Zahn2007}
Zahn O.,  Lidz A.,  McQuinn M.,  Dutta S.,  Hernquist L.,  Zaldarriaga M.,
  Furlanetto S.~R.,  2007, \mn@doi [The Astrophysical Journal]
  {10.1086/509597}, 654, 12

\makeatother
\end{thebibliography}



\appendix

\section{What the CNN sees}
\label{sec:What the CNN sees}

\begin{figure}
\includegraphics[width=\columnwidth]{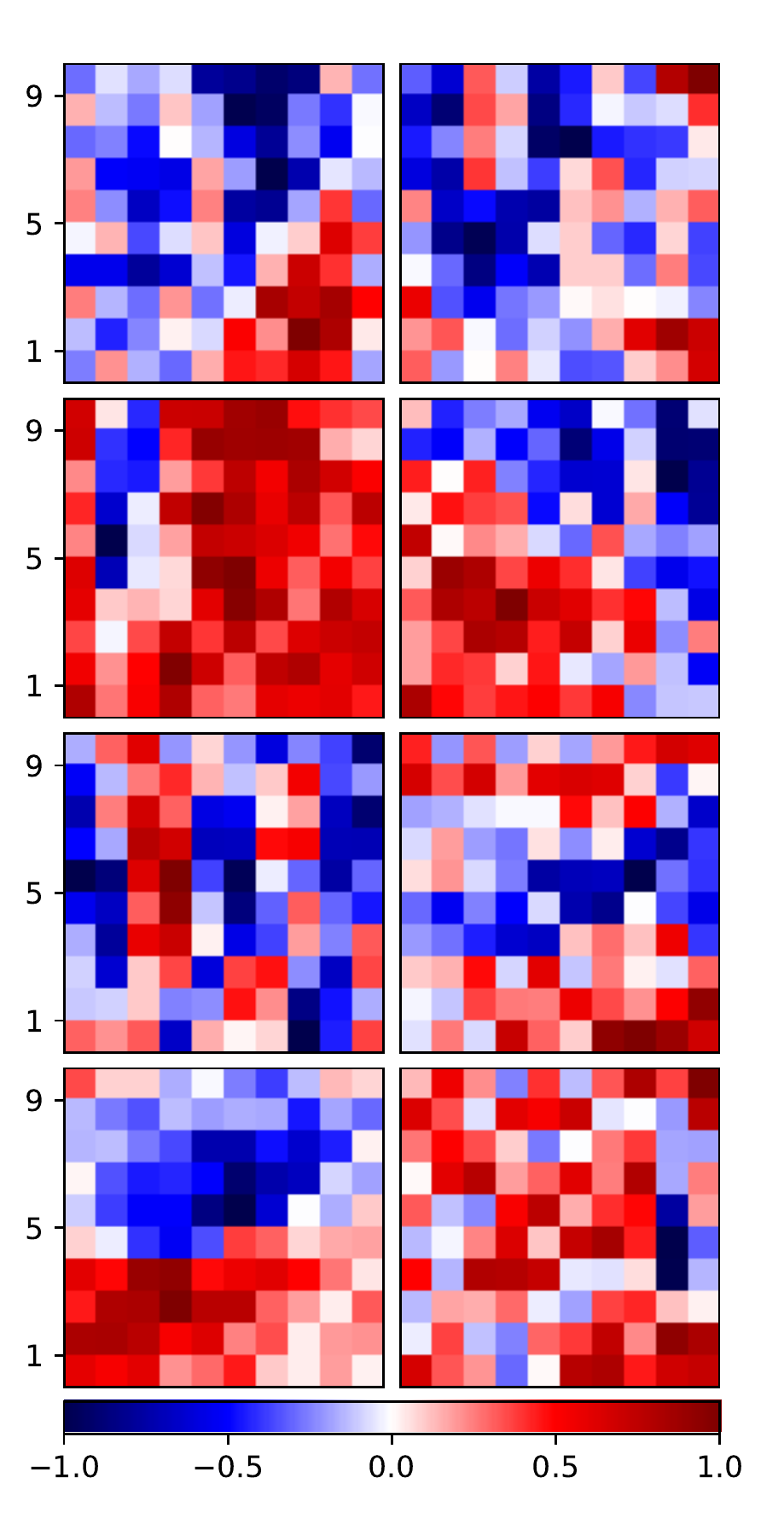} 
\caption{ Filters of the first convolutional layer after learning is completed (the color corresponds to a unit-less weight).  It is difficult to find obvious physical counterparts, though some could be interpreted as extractors of bubbles and/or connected web-like structures. Comparing to the convolution results shown in the following figure, the characteristic shapes and scales of these filters allow certain channels to specialize in particular cosmic epochs for some regions of parameter space.}
\label{fig:filters_conv1}
\end{figure}

\begin{figure*}
\includegraphics[width=\textwidth]{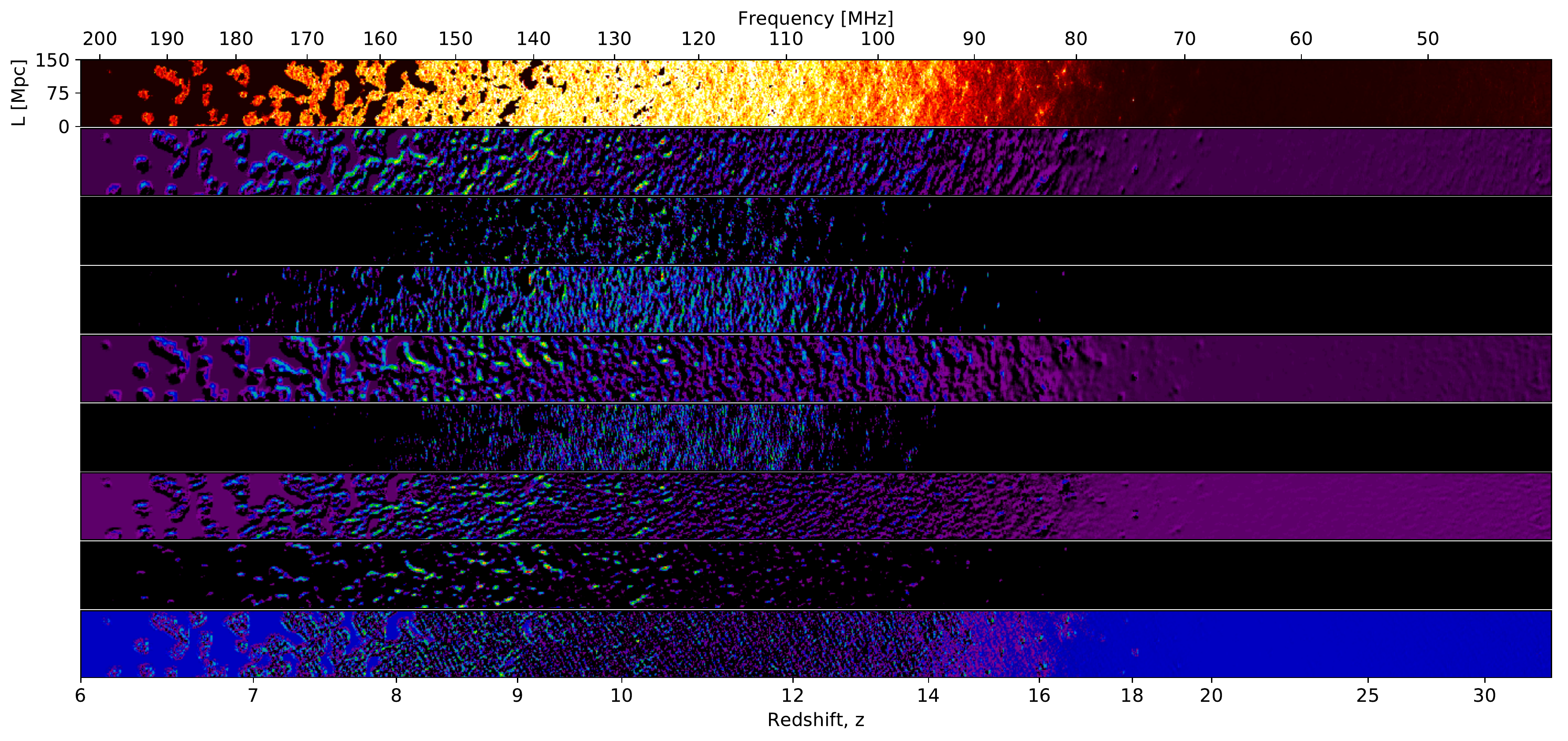} \\
\includegraphics[width=\textwidth]{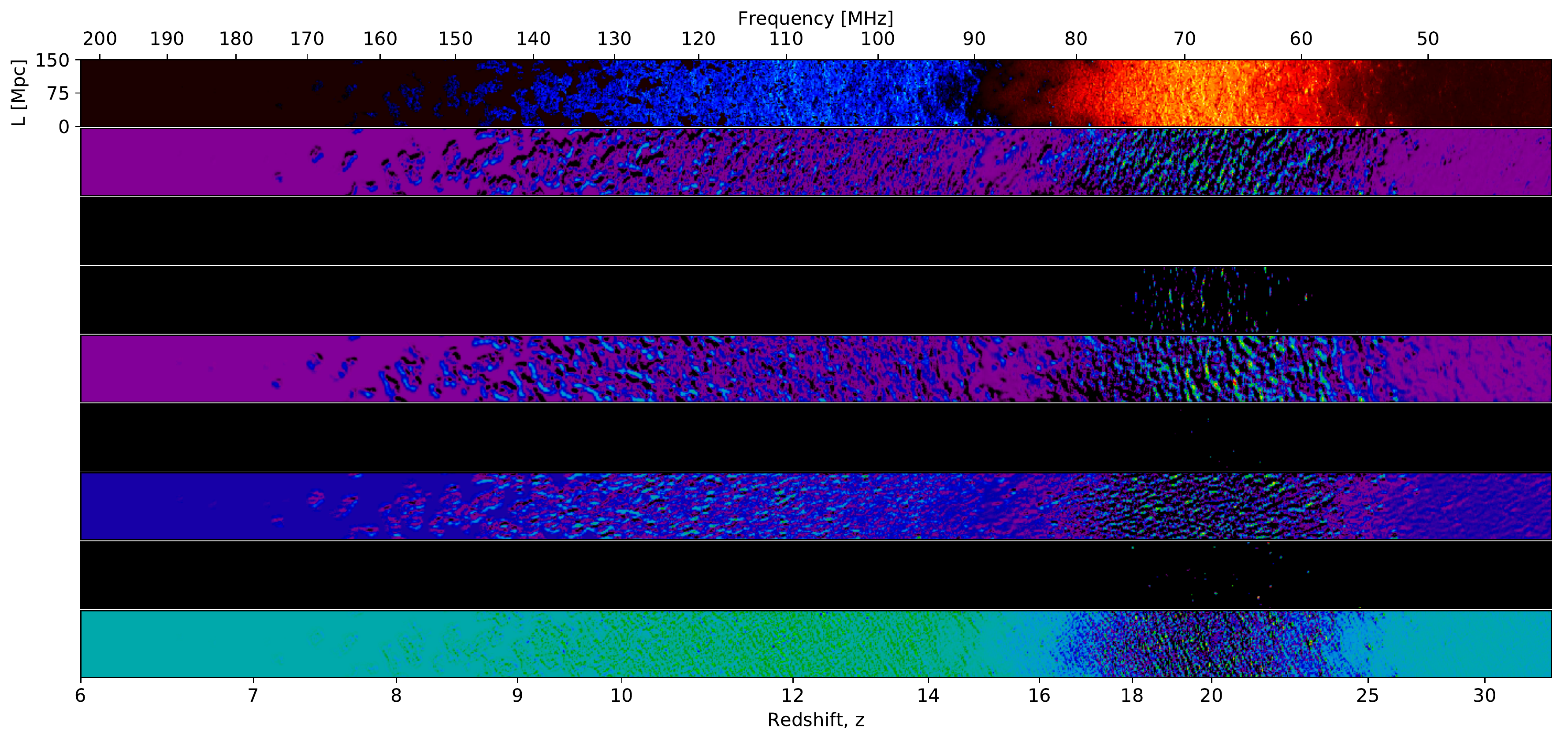}
\caption{ Response of the first convolutional layer to two different LC inputs, chosen to highlight different feature extraction, with [\Mzeta, \MTvir, \MLX, \MEo ]=[106, 5.2, 38.7, 0.89] and [30, 4.15, 40.5, 0.65], respectively. The first line of each panel is the inputed LC slice (with the same color bar as fig.\ref{fig:LC_faint_bright}). The following 8 lines are the response by each filters with the activation function (the first line of fig.\ref{fig:filters_conv1} correspond to the two first convolutions and so on).
}
\label{fig:conv1}
\end{figure*}

Oftentimes NNs are used as a \quotes{black boxes} because the interpretation of the internal weights is difficult. In the case of CNNs, the weights of the convolution layers correspond to filters. In order to peek inside the black box somewhat, in Figure \ref{fig:filters_conv1} we show the 8 filters of the first convolutional layer after the learning is completed. The weights are obviously not random anymore, and some shapes could be interpreted as extracting bubbles and web-like structures.  Other filters are difficult to interpret with human eyes. 

Another way to illustrate the CNN behavior is to look at the output of the convolution layers, after passing through the filters from the previous figure. In  Figure \ref{fig:conv1} we present two examples of the output of the first convolution layer, chosen to highlight different astrophysics and the resulting different network behaviors.  For both, the top panel corresponds to the inputted LC (the color bar is the same as in fig.\ref{fig:LC_faint_bright}).  The 8 following lines are the responses of the input by the 8 filters as shown in fig.\ref{fig:filters_conv1} (in left-to-right, top-to-bottom order).  The output contains the activation of the layer (the color is arbitrary). 

We can see that some of the channels (i.e. images produce by individual convolutions) seem to pick up reionization bubbles; others are more sensitive to heating structures, while others seem to also pick up density structures.  In the bottom example we can see that 3 channels are completely black, i.e. they do not include any information.  But those same channels in the top example contain information, picking up pre-reionization fluctuations.  This illustrates the fact that information takes a different path in the network, depending on the input values.

The interpretation of the first convolutional layer is possible because it is directly linked to the input data.  The interpretation of deeper layers become more challenging, as they involve increasing non-linearity, and include convolutions of convolved images.  Once the information reaches the fully connected layers, it is completely obscure to human interpretation. 

\bsp    
\label{lastpage}
\end{document}